\documentclass[9pt,twocolumn,twoside]{osajnl}
\DeclareMathOperator{\asinh}{asinh}
\journal{ao} 

\setboolean{shortarticle}{false}
\title{Analysis of performance and robustness against jitter of various search methods for acquiring optical links in space}
\author[1]{Gerald Hechenblaikner}
\affil[1]{Airbus Defence and Space, Claude-Dornier-Strasse, 88090 Immenstaad, Germany; Gerald.Hechenblaikner [at] airbus.com}

\begin{abstract}
We discuss various methods for acquiring optical links in space using a dedicated acquisition sensor.
Statistical models are developed and simple analytical equations derived that compare the performance between a single and dual spiral scan approach as well as between sequential and parallel acquisition of link chains. Simple derived analytical equations allow relating essential search parameters such as track width, variance of the uncertainty distribution, capture radius and scan speed to the probabilities of acquiring the links within a specific time.
We also assess the probability of failing to acquire a link due to beam jitter and derive a simple analytical model that allows determining the maximum tolerable jitter for a given beam overlap and required probability of success. All results are validated by Monte Carlo simulations and applied to the concrete example of the GRACE FO mission.
\end{abstract}

\setboolean{displaycopyright}{false}

\begin{document}

\maketitle

%%%%%%%%%%%%%%%%%%%%%%%%%%%%%%%%%%%%%%%%%%%%%%%%%%%%%%%%%%%%%
\section{Introduction}
\label{sec:intro}
\subsection{Optical Links in Space}
Optical links for inter-satellite communication have been established in Low-Earth-Orbit (LEO) between the  NFIRE and TerraSAR-X  satellites since 2007 \cite{fields2009nfire,smutny2008orbit} and achieve data rates of more than 5 Gbps over distances of up to 6000 km.
The European Data Relay System (EDRS) deploys improved terminals with respect to those used in LEO to establish links between geostationary satellites that provide routine data relay services to the Sentinel 1 and 2 satellites with over 1000 optical links established per month. Future extensions of the EDRS will add an additional geostationary node and bridge distances of up to 80 000 km between geostationary nodes\cite{hauschildt2019global}.

Aside from applications in communications, optical links are also used for precise interferometric measurements performed in space to map the earth gravitational field, such as done for the GRACE FO mission \cite{heinzel2017laser}\cite{schutze2014laser}\cite{abich2019orbit}, where optical links are established in between two spacecraft (SC) over a distance of approximately 200 km. These links allow measuring minute changes of the inter-SC distance on the order of nanometers, which provides information on the variations of the earth gravitational field.
In the original planning of this mission, the optical links were included as an experimental demonstration in addition to the microwave links that were based on proven technology. Owing to the superior performance of optical links, they are likely to be base-lined for use in future missions of this type as well.

Optical links also represent a key element for space-based gravitational wave detectors, such as the planned ESA/NASA LISA mission \cite{danzmann2003lisa}, where three spacecraft form an equilateral triangular constellation with an arm-length of approximately 2.5 million km across which interferometric measurements are performed. Similarly, the Chinese TAIJI mission uses a  triangular constellation with slightly larger arm-length of approximately 3 million km \cite{luo2020taiji}. The Japanese DECIGO mission \cite{sato2017status} aims to deploy a total of four different  triangular constellations around the sun, each constellation forming optical links across an arm-length of 1000 km. 

%%%%%%%%%%%%%%%%%%%%%%%%%%%%%%%%%%%%%%%%%%%%%%%%%%%%%%%%%%%%%%%%
\subsection{Link Acquisition Methods}
Before two SC can exchange information or perform interferometry over a bi-directional optical link, the latter has to be established first. For a pair of optical transceivers we shall refer to the link as "spatially acquired", if each transceiver (TC) detects the light transmitted from the respectively other transceiver. Following this step, one transceiver may lock its local laser onto the signal received from the remote transceiver by a phase-lock loop, which is referred to as "frequency acquisition". Such a two-step process is applied for the Laser Communication Terminals (LCTs) \cite{benzi2016optical} or for gravitational wave detectors in space \cite{cirillo2009control}.

Contrary to this approach, for the GRACE FO mission 5 degrees of freedom must be varied up to the point where all of them match simultaneously and the link is established\cite{mahrdt2014laser,wuchenich2014laser}. The degrees of freedom that are varied comprise the two pointing angles of each transmitted beam in addition to their relative frequency difference. The full 5-dimensional uncertainty space is covered by performing a two-dimensional angular scan in incremental steps with one beam and, for each increment, performing a full two-dimensional angular scan with the other beam. After a full four-dimensional coverage of all spatial scanning points, this sequence repeats for a different relative frequency offset between the two lasers, until a simultaneous match of all 5 degrees of freedom is detected, which is expected to take approximately 6 hours\cite{wuchenich2014laser}. As the optical instrument of GRACE FO  was initially only intended as an experimental demonstrator, the ensuing constraints did not allow using a dedicated acquisition sensor so that the instrument had to rely on the hardware required for science operations.

The LCTs used in the EDRS use the simplest possible acquisition sensor consisting of a quadrant photo-diode During the acquisition process one of the terminals, terminal 1, initiates a 2-dimensional search scan in angular space while the other terminal, terminal 2, detects any "hit" on its four quadrant sensor system and re-orients itself in accordance to the detected hits density distribution\cite{benzi2016optical,smutny2008orbit}. In the next step, the roles are swapped and terminal 2 scans while terminal 1 re-orients itself. Owing to the crude angular resolution of the quadrant photo-diode sensor, this process must be repeated iteratively a number of times until both terminals are aligned to within an accuracy of approximately 10 µrad and the tracking sensors take over.  

For gravitational wave detectors in space, the distances are typically very large, on the order of a million km, which requires large beam waists for transmission and results in very small beam divergence angles. Consequently, the beam jitter must be kept very small and the spacecraft accurately pointed in order to successfully acquire an optical link in a large uncertainty plane.  Therefore, a dedicated acquisition sensor with high angular resolution is used to support the acquisition process.
We will baseline such a sensor for the following discussions as it offers the simplest way of implementing the acquisition process as well as the best performance in terms of its total duration. Nonetheless, the general approach and results derived in the following sections do not depend on this assumption.

\subsection{Analytical models for acquisition performance}
Some of the first analyses on the acquisition process compared the characteristics of various search patterns (e.g., spiral, raster, Lissajous) and search methods (e.g., stare-scan, scan-scan) and assessed the effect of vibrations on the search time by simulation \cite{scheinfeild2000,scheinfeild2001}. 
The early analytical model of \cite{hindman2004} was used to assess the impact of the beam divergence angle on scan time. A more complex analytical model was developed in \cite{li2011analytical}, where a discretized spiral with individual dwell points was used, like the one introduced before in  \cite{scheinfeild2000}, to compute the mean search time for either a single scan or for a series of repetitive scans over the same region. This model was extended to include the effect of micro-vibrations on the mean search time in a very recent paper \cite{ma2021satellite}.
The impact of receiving terminal position on the probability of acquisition failure for a given level of beam jitter was studied with a complex analytical model in  \cite{friederichs2016vibration}.

One goal of this paper is to derive analytical expressions for the probability density function (PDF) of search times to acquire optical links in a constellation of satellites. We start by deriving the PDF for a single spiral scan which -to the best knowledge of the authors- has not been derived before. It is then used to obtain an analytical expression for the PDF of a novel “dual scan” method that utilizes a wide-field sensor in section 2. Building on these results, we obtain the PDFs of search times for a constellation of satellites, where 4 different methods are introduced and compared to another in relation to speed in section 3.
Another goal of this paper is to derive an analytical model to assess the impact of beam jitter on the probability of acquisition failure in section 4, which we then use to assess the robustness of the methods presented before. Unlike the models of \cite{ma2021satellite,friederichs2016vibration}, we include the effect of correlations between tracks, which -as we will see in section 4 of this paper- is quite important for a narrow pitch of the spiral track.
In summary, this paper aims to support the specification of high-level requirements for acquisition of a constellation of satellites by  deriving analytical models to assess some of the key performance aspects and to highlight the underlying parametric dependencies.

%%%%%%%%%%%%%%%%%%%%%%%%%%%%%%%%%%%%%%%%%%%%%%%%%%%%%%%%%%%%%
\section{Acquiring a bi-directional optical link}
In this section we investigate the general problem of spatially acquiring an optical link between two distant transceivers, referred to as TC-1 and TC-2.
\subsection{Beam detection}
\label{sec:beam_detection}
We assume that each transceiver transmits an optical beam towards the expected position of the other transceiver and that each transceiver can detect the beam transmitted towards it, if the beam is received by its entrance aperture within a certain field-of-view (Fov).
When a transceiver orients itself to point towards the expected position of the remote transceiver, the transmitted beam will generally miss the remote transceiver due to its overall pointing error. The latter generally comprises all sources of errors, including the attitude error of the transceiver platform (the spacecraft), the transceiver alignment error with respect to the platform (which dominates the uncertainty for GRACE FO\cite{wuchenich2014laser}), and the knowledge error of the position of the remote transceiver. These errors can be combined to yield an overall probability distribution of the pointing error of the transmitted beam which can generally be assumed to be an unbiased normal distribution with variance $\sigma^2$.

In general, the angular width of the transmitted beam only covers a small fraction of the uncertainty region within which the remote transceiver is expected to be. The remote transceiver is only able to detect the beam transmitted from the other transceiver if two conditions are fulfilled: (1) The remote transceiver is "hit" by the transmitted beam, i.e. it is located within the angular foot-print of the transmitted beam. (2) The incidence angle of the received beam is within the field-of-view (FoV) of the remote transceiver.
For illustration, Figure 1 shows the orientations of TC-1 with respect to the incident beam from TC-2. 
\begin{figure}
\centerline{\includegraphics[width=\columnwidth]{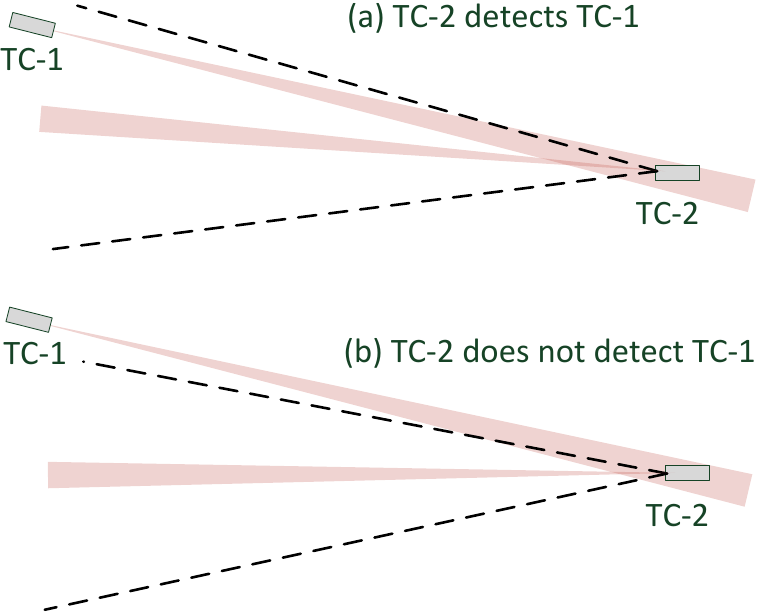}}
\caption  {Conditions for detecting the incident beam (a) TC-2 detects the beam from TC-1 as it is within the acquisition sensor FoV (dashed line) (b) TC-2 does not detect the beam from TC-1 as it is outside the acquisition sensor FoV . \label{fig:beam_detection}}
\end{figure}
%%%%%%%%%%%%%%%%%%%%%%%%%%%%%%%%%%%%%%%%%%%%%%%%%%%%%%%%%%%%
\subsection{The acquisition process}
\label{sec:acquisition_process}
\begin{figure}
\centerline{\includegraphics[width=\columnwidth]{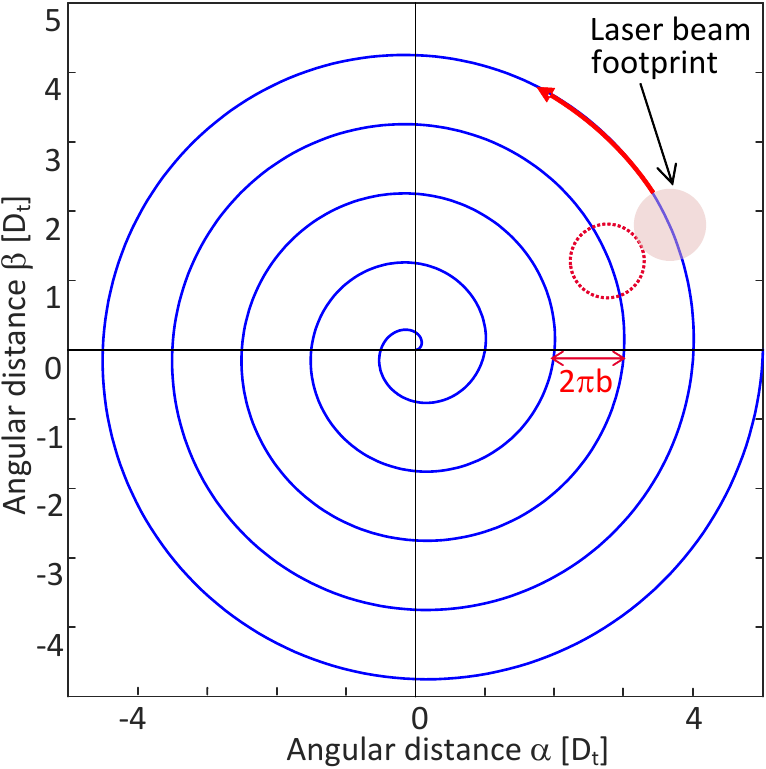}}
\caption  {Archimedean spiral used for scanning the uncertainty region. The red shaded area indicates the footprint (angular width) of the scanning scanning beam. \label{fig:spiral}}
\end{figure}

In the following we shall briefly summarize the acquisition process with a dedicated acquisition sensor as described in \cite{cirillo2009control}. We assume that each transceiver is equipped with an acquisition sensor of sufficiently large FoV to allow it detecting the incident beam from the other SC. The acquisition sensor is also assumed to be able to accurately resolve the position of the other SC in both angular degrees of freedom. 
One of the transceivers then starts scanning the uncertainty region in angular space where the other transceiver is expected to be, while the other transceiver is waiting. The scan follows a symmetric search pattern around the center of the uncertainty region. Such a pattern could be defined by an Archimedean spiral, as was introduced in \cite{cirillo2009control, mahrdt2014laser, luo2017possible}. It is also possible to choose a hexagonal or a rectangular pattern, or another symmetric pattern, but this reduces the efficiency of the proposed method. 
The Archimedean spiral is parameterized as follows
\begin{equation}
r=b\theta,
\label{eq:spiral}
\end{equation}
where $\theta$ denotes the spiral angle which increases by $2\pi$ with each spiral turn and $r$ denotes the spiral radius. The parameter $b$ defines incremental increase of spiral radius with each turn and relates to the track width $D_t$ (separation between spiral tracks, also referred  to as pitch) by the following relationship: $D_t=2\pi b$.
Figure \ref{fig:spiral} shows an Archimedean spiral and indicates the footprint of the scanning laser beam by the red-shaded area. The track-width of the spiral is chosen such that the footprint of the beam overlaps at two successive spiral turns which guarantees full coverage of the uncertainty area.
As soon as the waiting transceiver has detected the incident beam of the scanning transceiver, it re-orients itself (or more generally its transmitted beam and acquisition sensor) to point towards the angular position of the received beam which was resolved by the acquisition sensor. At this point the scanning transceiver then also detects a signal on its acquisition sensor, stops its search spiral, and re-orients itself accordingly. At this point a bi-directional link between both transceivers has been spatially acquired.
%%%%%%%%%%%%%%%%%%%%%%%%%%%%%%%%%%%%%%%%%%%%%%%%%%%%%%%%%%%%%%%%%%%%%%%%%%%%%%%%%%%
\subsection{Probability distribution of a spiral search}
\label{sec:ProbDistr}
\begin{figure}
\centerline{\includegraphics[width=\columnwidth]{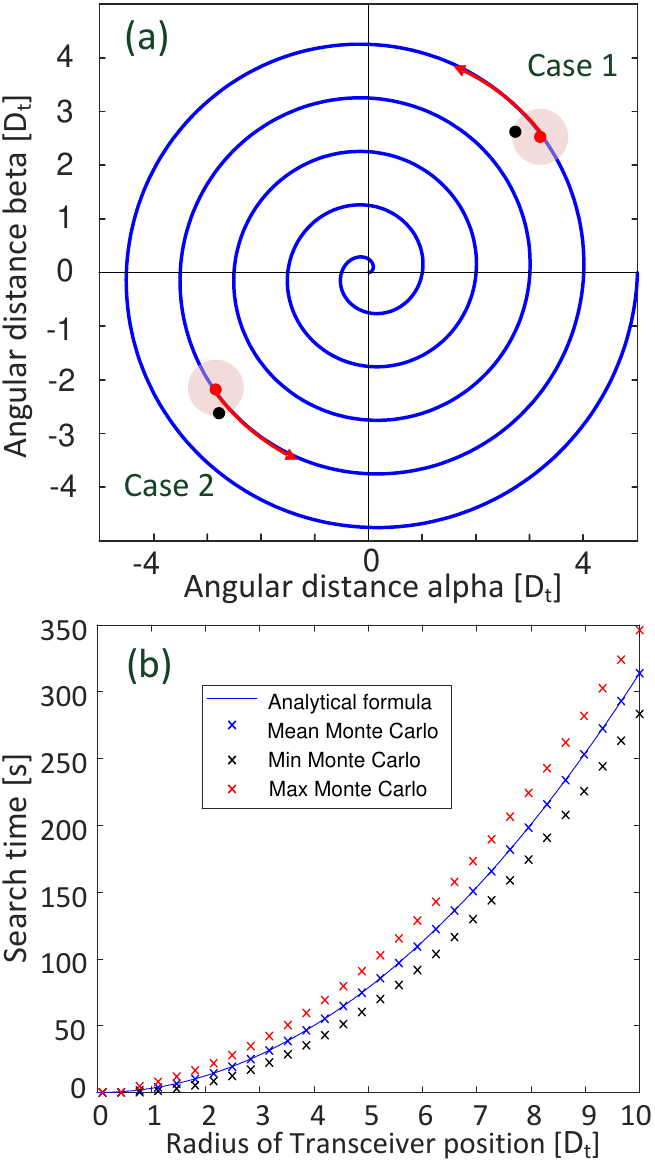}}
\caption  {(a) Two possible positions of the remote TC within the uncertainty region for a given radius R. (b) Distribution of search times to detect the remote TC located at fixed radius R\label{fig:radial_position}}
\end{figure}

From Eq.\ref{eq:spiral} we can calculate the total length of the spiral by integrating along the spiral path. It is given by
\begin{equation}
L=1/2\left(r\sqrt{1+\left(r/b\right)^2}+b\asinh r/b\right)
\label{eq:total_spiral_length}
\end{equation}
In the limit of $r\gg b$, Eq. \ref{eq:total_spiral_length} simplifies to $L\approx r^2/(2b)$. Assuming constant scan speed, the time $T$ it takes to reach a certain spiral radius is simply found by dividing $L$ by the scan speed $\gamma$.
One can also arrive at this result by solving the equations of motion for the assumption that the angular velocity of the scanning beam is constant, i.e. $\gamma= v_{\theta}=r\dot{\theta}$.
Using the spiral equation \ref{eq:spiral} then yields the differential equation $r\dot{r}=b\gamma$ from above and which was already introduced in \cite{cirillo2009control}.
\begin{equation}
r(t)=\sqrt{2b\gamma t} \rightarrow t=\frac{r^2}{2b\gamma}
\label{eq:search_time}
\end{equation}
We note that Eq. \ref{eq:search_time} accurately predicts the time it takes to "hit" the remote TC if it is located exactly at some point on the search spiral. This situation will generally not be the case, as it will be located somewhere on either side of the spiral track but not exactly on it. Figure \ref{fig:radial_position}a shows two cases of potential locations of the remote TC, indicated by the black filled circle, within the uncertainty area for the same radial distance $R$. In case 1, the search time will be longer than predicted by Eq.\ref{eq:search_time} because the scanning beam, indicated by the red filled circle, has reached a larger radial distance than the TC when it is being detected. In case 2, it is the other way round. 
However, the average over all search times for a fixed radial position agrees exactly with the prediction of Eq.\ref{eq:search_time}, as can be seen in Fig.\ref{fig:radial_position}b, where the analytical formula (blue solid line) is plotted together with the results of Monte Carlo simulations for randomly chosen locations of the TC within the uncertainty plane, assuming a scan speed of $\gamma=D_t$per second. The figure plots the average (blue crosses) as well as the minimum (black crosses) and maximum (red crosses) search times for a given radial distance. The fact that the statistical average can be exactly expressed by Eq.\ref{eq:search_time} is an important property that we will exploit to derive the probability density function of search times.  

In the following we shall assume that the uncertainty of the location of the remote transceiver TC-1 in the uncertainty plane seen by transceiver TC-2 is represented by two independent random variables, $\alpha_R$ and $\beta_R$, which are assumed to be normally distributed with zero mean and the same variance  $\sigma^2_{\alpha}=\sigma^2_{\beta}=\sigma^2$ . Owing to the radial symmetry, the associated joint probability density function $PDF(\alpha,\beta)$ for the angles $\alpha$ and $\beta$ in the uncertainty plane can be converted into a probability density function $PDF(r)$ for a random variable $R=\sqrt{\alpha_R^2+\beta_R^2}$ which is associated with the radial distance $r$ in the uncertainty plane\cite{li2011analytical}:
\begin{equation}
PDF(\alpha,\beta)d\alpha d\beta=\frac{d\alpha d\beta}{2\pi\sigma^2}e^{-\frac{\alpha^2+\beta^2}{2\sigma^2}}=\frac{r dr}{\sigma^2}e^{-\frac{r^2}{2\sigma^2}}=PDF(r)dr
\end{equation}
The random variable $T$ that is associated with the search time relates to the random variable $R$ that is associated with the radial position of the remote TC through Eq. \ref{eq:search_time}: $T=R^2/(2b\gamma)$. The associated probability density function $PDF(t)$ for $T$ can then be found through an integral transformation of $PDF(r)$ using the relation $r dr=b\gamma dt$:
\begin{equation}
PDF(r) dr=\frac{r dr}{\sigma^2}e^{-\frac{r^2}{2\sigma^2}}=\frac{b\gamma}{\sigma^2}e^{-\frac{b\gamma}{\sigma^2}t}dt=PDF(t)dt
\label{eq:PDF_derived}
\end{equation}
The same result can be obtained immediately when noting that the random variable $T$ must be proportional to the $\chi^2$ distribution with 2 degrees of freedom which is an exponential function. This is because $T$ is proportional to the sum of the squares of two standard normal variates defining the random variables $\alpha_R$ and $\beta_R$ via the relation $T\propto R^2\propto \alpha_R^2+\beta_R^2$.

\begin{figure}
\centerline{\includegraphics[width=\columnwidth]{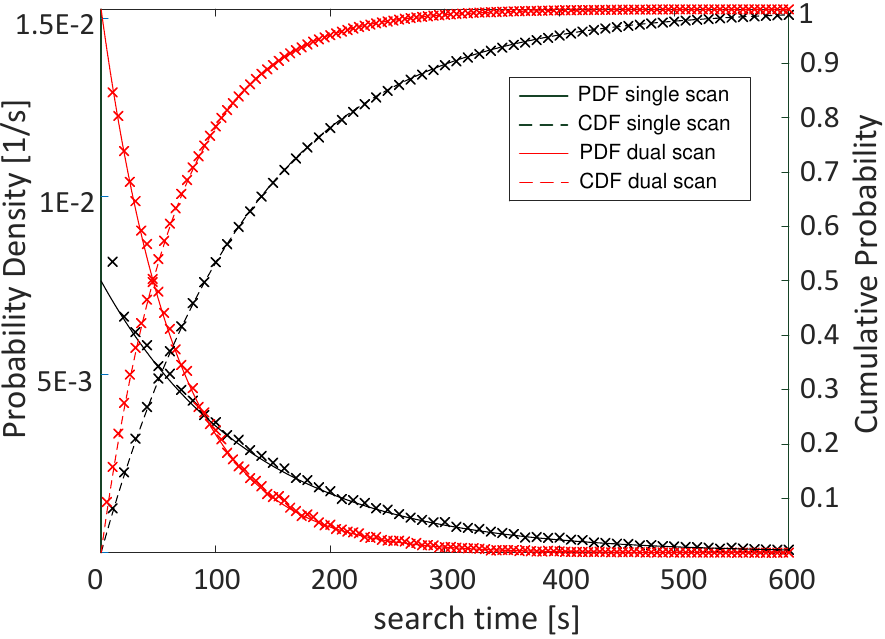}}
\caption  {Probability density functions (solid lines, referring to left ordinate) and cumulative distribution functions (dashed lines, referring to right ordinate) are plotted against the search time. The black curves refer to a single search scan, the red curves to a dual scan where both transceivers are scanning simultaneously. The results of Monte Carlo simulations are indicated by black and red crosses, respectively. Simulation assumptions: $D_t$= 270 $\mu$rad, $\sigma$=872 $\mu$rad, $\gamma$=$D_t$/2 per second; $1/\lambda=131 s$. \label{fig:exponential_functions}}
\end{figure}

Expressing Eq.\ref{eq:PDF_derived} in terms of the fundamental scaling parameter $\lambda=b\gamma/\sigma^2$, we obtain the PDF and after integration the cumulative distribution function (CDF) of the search process which we refer to as $f_{\lambda(t)}$ and $F_{\lambda}(t)$, respectively.
\begin{eqnarray}
f_{\lambda}(t)&=&\lambda e^{-\lambda t}\label{eq:PDF_lambda}\\
F_{\lambda}(t)&=&1-e^{-\lambda t}\label{eq:CDF_lambda}
\end{eqnarray}
We shall now discuss these results for the specific case of the GRACE FO mission. It is conceivable that in a future successor of the current mission an optical instrument will be base-lined and equipped with an acquisition sensor. From the system parameters specified in  \cite{heinzel2017laser} we can determine the beam divergence angle $\theta_{div}=\lambda/(\pi w_0)= 135\, \mu$rad. We then choose the track width $D_t$ of the search spiral to be twice as large in order to obtain full coverage of the spiral area, yielding a level of received power between $190\,\text{pW}$ in the beam center and $27\,\text{pW}$ at the edge of the track, which is sufficient for detection by sensitive CCD arrays under the assumption of limited levels of background stray light.
The specified "uncertainty region" radius of $R_{uc}$=3 mrad \cite{wuchenich2014laser} could be interpreted as the radius in the uncertainty plane within which the remote SC is located with a probability of 99.73\% ("3$\sigma$ value of the normal distribution). In the 2d plane this corresponds to a standard deviation of the two angles $\alpha$ and $\beta$  that is given by $\sigma=R_{uc}/3.44=872\,\mu \text{rad}$. 
The choice of spiral scan speed primarily depends on the limitations of the actuator but also on the required detector integration time and the beam jitter.
We shall assume it is given by half of the track width per second and obtain $\gamma=135\, \mu\text{rad/s}$. We then find for the characteristic parameter $\lambda=7.63E-3$.
From Eqs.\ref{eq:PDF_lambda},\ref{eq:CDF_lambda} it is useful to derive the following relations for the mean search time $\overline{T}$, and the time $T_{p}$ within which the search is concluded for a specified probability of success $P$:
\begin{eqnarray}
\overline{T}&=&1/\lambda=\frac{\sigma^2}{b\gamma}\label{eq:mean_time_exp}\\
T_{p}&=&\frac{-\ln\left(1-P\right)}{\lambda}=\frac{\sigma^2}{b\gamma}\ln\left(1-P\right)\label{eq:P_time_exp}
\end{eqnarray}
We find for the assumed parameters of the GRACE FO mission given above that $\overline{T}=131 s$ and the time within which the search is concluded with 99.73\% probability $T_{99.73}=775 s$.
In order to confirm the validity of Eqs.\ref{eq:PDF_lambda},\ref{eq:CDF_lambda} we performed a total of 1E5 Monte Carlo simulations where the laser beam is moved along an Archimedean spiral up to the point where a randomly positioned SC falls within its capture radius.
The results are given in Fig.\ref{fig:exponential_functions} by the black curves and associated black crosses, referring to the analytical predictions of Eqs.\ref{eq:PDF_lambda},\ref{eq:CDF_lambda} and the Monte Carlo results, respectively. There is very good agreement between prediction and simulation. The simulation data point for the shortest time deviates from the prediction,  as it includes the small probability of finding the SC right at the centre of the uncertainty plane at t=0, but this has no noticeable impact on the general results.

%%%%%%%%%%%%%%%%%%%%%%%%%%%%%%%%%%%%%%%%%%%%%%%%%%%%%%%%%%%%%%%%%%%%%%
\subsection{Dual scan acquisition method}
\label{sec:dual_scan}
In this section we discuss an alternative method to the acquisition which has advantages in terms of speed and is more robust against failure due to jitter (discussed in section \ref{sec:beam_jitter}). 
Instead of a single receiver performing a scan while the other is waiting to detect the transmitted beam (as discussed in \cite{cirillo2009control}), both transceivers  perform a spiral search scan simultaneously. This is similar to the approach taken in GRACE FO \cite{wuchenich2014laser} or an alternative concept \cite{luo2017possible}, but instead of aligning all 5 degrees of freedom simultaneously, a wide-field acquisition sensor is used. Once either of the two transceivers detects the other, it stops the scan and re-orients itself in the according direction. Then the other scanning transceiver likewise detects the transmitted beam and re-orients itself to point into the direction of the received beam, at which point the spatial acquisition has been concluded. \\
It should be noted that this process requires a FoV that is twice as large as the one required if only one transceiver is scanning. This is because a transceiver may - in a worst case assumption- receive a beam from the other transceiver when it is itself scanning at the outer fringe of the uncertainty region. Considering that the received beam itself could also come from the outer fringe of the uncertainty region, and assuming that the acquisition sensor FoV is centered around the direction of the transmitted beam, this requires a twice as large FoV to guarantee detection for all possible cases. Additionally, there will be a slight increase of the sport size of the received beam on the acquisition sensor, because the image will blur out slightly due to the scanning motion of the receiving transceiver. However, this increase is small because the scanning motion only covers a tiny fraction of the acquisition sensor field-size during the integration time of the detector.\\
In the following we want derive the PDF and CDF for the search times if a dual scan approach is used.
We  assume that the two independent random variables $T_1$ and $T_2$ represent the times it would take for the scanning transceiver TC-1 to hit TC-2 and for the scanning transceiver TC-2 to hit TC-1, respectively. They are both assumed to be exponentially distributed according to Eq.\ref{eq:PDF_lambda}. As detailed before, the search is concluded as soon as the transmitted beam of either transceiver is incident on the other transceiver. The random variable for the acquisition time $T$ can therefore be represented as the minimum of the two random variables  $T_1$ and $T_2$:
\begin{displaymath}
T=\min\left(T_1,T_2\right)
\end{displaymath}
We then find for the cumulative distribution function $F_{min}(t)$ of $T$:
\begin{eqnarray}
F_{min}(t)&=&\text{Prob}(T\leq t)=1-\text{Prob}(T>t)\nonumber\\
&=&1-\text{Prob}(T_1>t)\times\text{Prob}(T_2>t)\label{eq:P_T1T2}
\end{eqnarray}
Using Eq.\ref{eq:CDF_lambda}, we find for the probability that the random variable $T_1$ is larger than a given time $t$
\begin{equation}
\text{Prob}(T_1>t)=1-\text{Prob}(T_1\leq t)=e^{-\lambda t},\label{eq:T1_larger_t}
\end{equation}
and like wise for $T_2$. Inserting Eq.\ref{eq:T1_larger_t} into Eq.\ref{eq:P_T1T2} we obtain for the CDF and PDF of the dual search scan, respectively:
\begin{eqnarray}
F_{min}(t)&=&\text{Prob}(T\leq t)=1-e^{-2\lambda t}\label{eq:CDF_dual}\\
f_{min}(t)&=&\frac{d}{dt}F_{min}(t)=2\lambda e^{-2\lambda t}\label{eq:PDF_dual}
\end{eqnarray}
We conclude that a dual scan methods also implies an exponential distribution of search times, but with an exponent that is twice as large as the one for the single scan method.
Consequently, Eqs.\ref{eq:mean_time_exp},\ref{eq:P_time_exp} can be used upon substitution of $\lambda$ by $2\lambda$, yielding search times that are half as long as for the single scan method. For example, in the case of the GRACE FO acquisition parameters used to generate Fig.\ref{fig:exponential_functions}, we obtain a mean search time of 66 seconds.
There are more advantages of a dual scan performed by both transceivers compared to a single scan performed by only one which we discuss in section \ref{sec:beam_jitter}.

The red curves of Fig.\ref{fig:exponential_functions} give a comparison between the predictions of Eqs.\ref{eq:CDF_dual},\ref{eq:PDF_dual} and the results of Monte Carlo simulations, where the interaction between two transceivers was simulated and a statistics compiled. There is excellent agreement between the two, confirming the validity of the approach. 
It should be noted that the crucial assumption was made that the two random variables are independent in the derivation of Eq.\ref{eq:PDF_dual}, which is equivalent with the assumption that the random variables $\alpha_R$ and $\beta_R$, defining the random location of one transceiver in the uncertainty plane seen by the other transceiver, are uncorrelated between the two transceivers. 
If they are not, the efficiency of the dual scan method is decreased and, in the case of full correlation, will be the exactly the same as for the single scan method only.
In particular, such a correlation exists, if the knowledge error of the relative transceiver positions (which is required for pointing into the right direction) is dominant in the overall pointing error.
However, in many applications this is not the case. As an example, in the case of GRACE FO, the random variables can be considered as uncorrelated, as the primary factors entering the overall uncertainty are the orientation of the transceiver unit with respect to the SC platform  and the attitude error of the platform itself \cite{wuchenich2014laser}.

%%%%%%%%%%%%%%%%%%%%%%%%%%%%%%%%%%%%%%%%%%%%%%%%%%%%%%%%%%%%%%%%%%%%%%%%
\section{Acquisition of Link Chains}
\begin{figure}
\centerline{\includegraphics[width=\columnwidth]{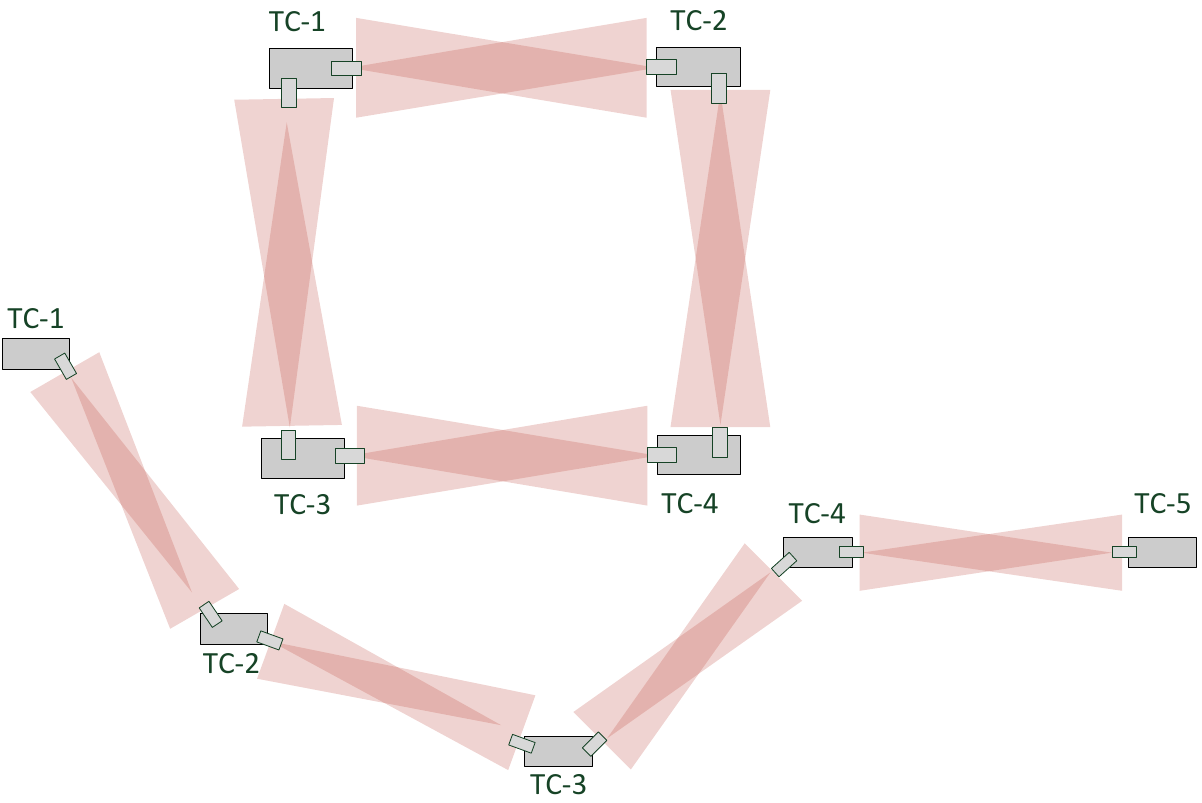}}
\caption  {Examples of link chains: A closed link chain with 4 nodes and an open link chain with 5 nodes. \label{fig:link_chains}}
\end{figure}

In this section we discuss the extension of the concept of a link between two transceivers to the concept of a link chain, where multiple transceivers are connected via optical links.
Such a chain can be used for optical communication in a network, such as the future extension of the EDRS \cite{hauschildt2019global}, or for a satellite formation to monitor the Earth or to detect gravitational waves.
There are two types of transceiver chains: open transceiver chains and closed ones, as shown in Fig.\ref{fig:link_chains}. The acquisition scheme for an open transceiver chain with N+1 nodes is the same as for a closed chain with N nodes, because in both cases a total of N links have to established between two partners.
The acquisition of a link chain can either follow a sequential method, where one link between a transceiver pair is built up after the other, or it can follow a parallel method, where each transceiver pair simultaneously with all others aims to establish a link.
%%%%%%%%%%%%%%%%%%%%%%%%%%%%%%%%%%%%%%%%%%%%%%%%%%%%%%%%%%%%%%%%%%%%%%%%%%%%%%%%%%%%
\subsection{Sequential Acquisition}
We try to find the total time to acquire all links in a chain on N links, neglecting the generally short and deterministic times for re-orientation of the transceivers as part of the acquisition process. The total time is then represented by a random variable $T$ that is given by the sum of $N$ independent and exponentially distributed random variables, each representing the search time for one particular link.
\begin{equation}
T=T_1+T_2+T_3+...+T_N
\end{equation}
The PDF of $T$ can be found quite easily by considering the relationship between the exponential distribution and the Poissonian distribution $P(k,\mu)$ which defines the probabilities of $k$ events occurring for a given average probability $\mu$. Considering that in our problem the average probability is given by the mean rate of successful searches within a given time we set $\mu=\lambda t$ and obtain
\begin{equation}
P(k,\lambda t)=\frac{e^{-\lambda t}(\lambda t)^k}{k!}.
\end{equation}
The probability of acquiring all $N$ links within a certain time $t$ is equal to the probability of concluding at least $N$ successful searches within that time. Denoting the number of successful searches $N_{s}$, we find
\begin{equation}
\text{Prob}(N_{s}\geq N)=1-\text{Prob}(N_{s}<N)=1-\sum_{k=0}^{N-1} P(k,\lambda t).
\label{eq:sequential_search}
\end{equation}
The probability defined in Eq.\ref{eq:sequential_search} specifies the cumulative distribution function $F_{s}(t)$ for the sequential acquisition times of a chain of $N$ links, from which the probability density function $f_s(t)$ is obtained by differentiation:   
\begin{equation}
f_s(t)=\frac{\lambda^N t^{N-1}e^{-\lambda t}}{(N-1)!}
\label{eq:Erlang}
\end{equation}
It should be noted that this PDF is identical to the Erlang distribution $f_{Erl}(t,\lambda,N)$ which is well known in the field of traffic engineering and stochastic processes. 

\begin{figure}
\centerline{\includegraphics[width=\columnwidth]{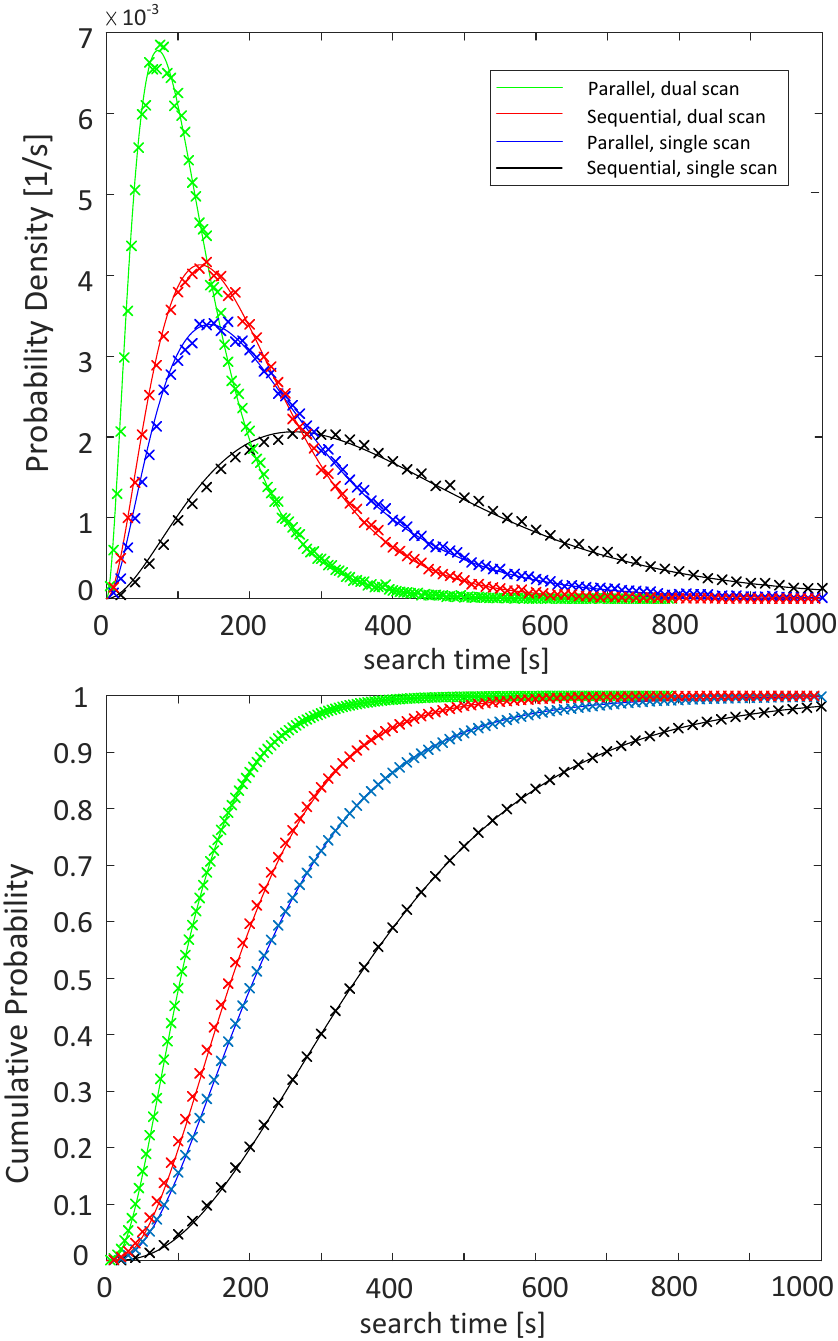}}
\caption  {Probability densities (a) and cumulative probabilities (b) for various formation acquisition options, from left to right curve: 1. parallel acquisition with dual scan (green), 2. sequential acquisition with dual scan (red), 3. parallel acquisition with single scan (blue), 4. sequential acquisition with single scan (black). Analytical predictions are denoted by solid lines, Monte Carlo results (1E5 simulation runs) by crosses. Simulation assumptions: $D_t$= 270 $\mu$rad, $\sigma$=872 $\mu$rad, $\gamma$=$D_t$/2 per second; $1/\lambda=131 s$. \label{fig:acquisition_options}}
\end{figure}

As an example, we plot in Fig.\ref{fig:acquisition_options} the CDF and PDF for the sequential acquisition of three links  (black solid lines) which represent a closed link chain between nodes in a triangular formation, as will be used for gravitational wave detectors in space \cite{danzmann2003lisa,luo2020taiji,kawamura2019space}. Triangular formations are also conceivable for future earth gravity mapping missions, such as a successor to GRACE FO, in order to extend the spatial resolution. The parameters used for generating the plots are discussed in section 2\ref{sec:ProbDistr} and summarized in the figure caption. 
Figure \ref{fig:acquisition_options} also plots the PDF and CDF for the case that each link is acquired using a dual scan approach, given by the blue lines, which are computed by substituting the fundamental scaling parameter $\lambda$ for $2\lambda$ in Eqs.\ref{eq:sequential_search} and \ref{eq:Erlang}. The results of Monte Carlo simulations that represent the full sequential acquisition process are indicated by the respective crosses in the figure and found to be in very good agreement with the analytical predictions.
%%%%%%%%%%%%%%%%%%%%%%%%%%%%%%%%%%%%%%%%%%%%%%%%%%%%%%%%%%%%%%%%%%%%%%%%%%%%%%%%%%%%%%%%%%%%%%%%%%%%%%%%%%%%%%%%%%%%%%%%%%%%%%%%%%%%%%%%%%%%%%
\subsection{Parallel Acquisition}
The links of a chain can also be acquired in parallel, where each pair of transceiver units comprising a certain link aims to establish the link in parallel with the pairs of all other links.
Each pair in itself can either follow a single scan method, where only one of the transceiver units is scanning the uncertainty region of the other, or it can follow a dual scan method, where both of the transceiver units are scanning at the same time.
The acquisition of the full chain is complete when the last link between a pair of transceiver units has been established.
The total acquisition time can therefore be represented by random variable $T$ which is defined as the maximum of $N$ independent random variables $T_1,T_2,...,T_N$, each representing the time it requires to acquire one of the links in the chain.
\begin{equation}
T=\max(T_1,T_2,T_3,..,T_N)
\label{eq:T_max}
\end{equation}
The random variables for the single link times are exponentially distributed with a scaling factor which is either given by $\lambda$ for the single scan method (see Eq.\ref{eq:PDF_lambda}) or  $2\lambda$ for the dual scan method (see Eq.\ref{eq:PDF_dual}).
We find the cumulative distribution function of the total acquisition time $T$, denoted as $F_{p}(t)$, as follows:
\begin{eqnarray}
F_{p}(t)&=&\text{Prob}(T\leq t)=\text{Prob}(\max(T_1,T_2,...,T_N)\leq t)\nonumber\\
&=&\text{Prob}(T_1\leq t)\times...\times\text{Prob}(T_N\leq t)\label{eq:parallel_search}
\end{eqnarray}
We note that $\text{Prob}(T_i\leq t)$ for $i=1...N$ is given by the cumulative distribution function of a single link, defined in Eq.\ref{eq:CDF_lambda} for the single scan method and in Eq.\ref{eq:CDF_dual} for the dual scan method. Assuming single scan and inserting it into Eq.\ref{eq:parallel_search}, we obtain $F_p(t)$ and after differentiation the probability density function $f_p(t)$:
\begin{eqnarray}
F_p(t)&=&\left(1-e^{-\lambda t}\right)^N\label{eq:CDF_parallel}\\
f_p(t)&=&N\lambda e^{-\lambda t}\left(1-e^{-\lambda t}\right)^{N-1}\label{eq:PDF_parallel}
\end{eqnarray}
Note that in case a dual scan method is applied, the fundamental scaling parameter $\lambda$ of Eqs.\ref{eq:CDF_parallel},\ref{eq:PDF_parallel} must be substituted for $2\lambda$.
As an example, we plot in Fig.\ref{fig:acquisition_options} the CDF and PDF for the parallel acquisition of three links  for the single scan method (red solid lines) and the dual scan method (green solid lines). 
Once again we used the parameters which are discussed iin section 2\ref{sec:ProbDistr} in the context of the GRACE FO mission and which are summarized in the figure caption. 
The analytical predictions agree very well with the results of Monte Carlo simulations that represent the full parallel acquisition process, indicated by the respective crosses in the figure.

So far we have looked at a total of four different options for how to acquire a link chain, the predictions of which are summarized in the plots of Fig.\ref{fig:acquisition_options}.
Note that all analytical curves scale with the fundamental scaling parameter $\lambda$ (see Eq.\ref{eq:PDF_lambda}) which is given by $1/\lambda=131 s$ for the chosen parameters.  This fact  can be used to adjust the plots to any given set of parameters for other acquisition scenarios.
Out of these four options, the option of using a dual scan method for each link and acquiring all of the links in parallel is clearly superior to the other options in terms of efficiency. However, it requires a larger FoV of the acquisition sensor and has the highest complexity.
The option of using a single scan method for each link and acquiring all of the links sequentially, is the slowest, but requires a smaller acquisition sensor FoV and is the least complex.
The acquisition times of the 4 options are summarized in Table \ref{tab:acquisition_times} which lists the mean, the median and the times within which the formation is acquired for a specified probability $P$ of success.  For convenience, the probabilities $P$ are expressed in units of standard deviation $\sigma_n$ of a normal distribution (for example, 3 $\sigma_n$ stands for 99.73 \% probability of success).
%%%%%%%%%%%%%%%%%%%%%%%%%%%%%%%%%%%%%%%%%%%%%%%%%%%%%%%%%%
\begin{table}[h]
\centering
\caption{\label{tab:acquisition_times}
Acquisition times with respect to mean, median and specified probabilities $P$ for the four options: 1. parallel acquisition with dual scan, 2. sequential acquisition with dual scan, 3. parallel acquisition with single scan, 4. sequential acquisition with single scan. Parameters as for Fig.\ref{fig:acquisition_options}}
\begin{tabular}{crrrr}
\hline
\textrm{}& \textrm{Opt. 1 [s]}& \textrm{Opt. 2 [s]}& \textrm{Opt. 3 [s]}& \textrm{Opt. 4 [s]}\\
\hline
mean & 120 & 197 & 240 & 393\\
median & 103 &175 & 207 & 351\\
1 $\sigma_n$  & 139 & 231& 278 & 461\\
2 $\sigma_n$ & 274 & 421 & 547 & 842\\
3 $\sigma_n$ & 460 & 657 & 919 & 1315\\
4 $\sigma_n$ & 706 & 947 & 1411 & 1895\\
5 $\sigma_n$ & 1014 & 1294 & 2028 & 2588\\
\end{tabular}
\end{table}

%%%%%%%%%%%%%%%%%%%%%%%%%%%%%%%%%%%%%%
\section{The Impact of Beam Jitter}
\label{sec:beam_jitter}
So far we have assumed that the transmitted beam is centred perfectly on the trajectory of the search spiral while the transceiver is scanning the uncertainty region. In this section we will investigate more closely what is happening under more realistic conditions, namely that the scanning beam is jittering around the ideal trajectory.
\subsection{The jitter model}
\begin{figure}
\centerline{\includegraphics[width=\columnwidth]{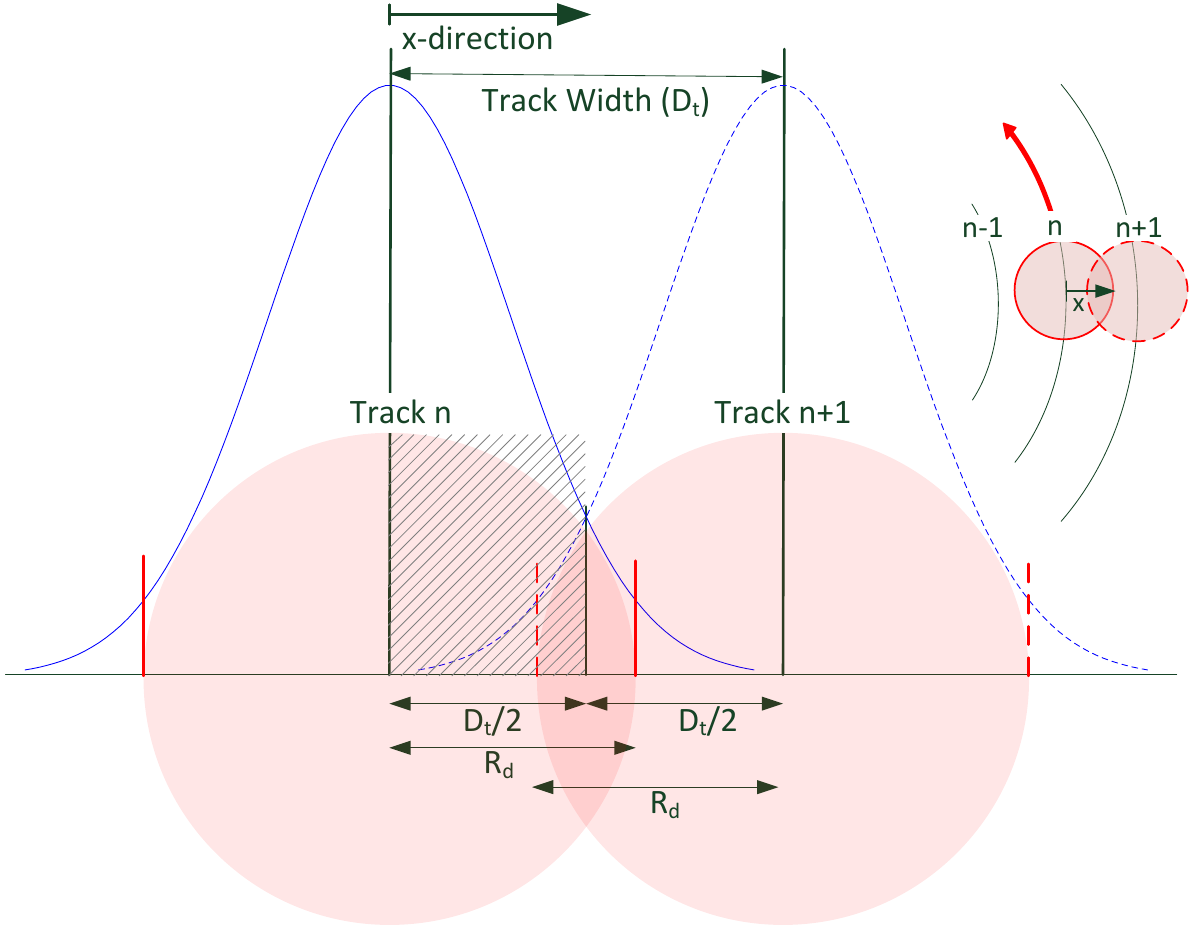}}
\caption  {The beam profile of the scanning beam for two adjacent tracks, $n$ and $n+1$, of the search spiral. Laser beam profiles shown as blue curves. The radius of detection ($R_d$) is defined to be at the beam waist,  Track Width=$D_t$, Radius of Detection=$R_d$  \label{fig:overlapping_beams}}
\end{figure}

Figure \ref{fig:overlapping_beams} defines the relevant properties needed for the following discussion.
It shows a section of the spiral plane along a radial vector $R_{TC}$ between the spiral center and the position of the transceiver in the spiral plane (see zoom-out in top right of figure). We define the origin of the "x-axis" of this plot to lie on spiral track  $n$ and to point outwards in radial direction. The remote transceiver is assumed to be located somewhere in between spiral track $n$ and half the distance to spiral track $n+1$, indicated by the hatched area in Fig \ref{fig:overlapping_beams}. The beam capture radii are represented by the red-shaded area and the laser beam intensity profiles along the radial direction for track $n$ and track $n+1$ are represented by the solid and dashed blue curves, respectively.
The acquisition sensor of the remote transceiver is able to detect the scanning beam if the average intensity during integration is higher than a specified limit that defines the radius of detection ($R_d$) of the beam, which we assume to coincide with the beam waist. The track width ($D_t$) of the search spiral is chosen such that $D_t/2$ is smaller than $R_d$ so that the beams overlap by an amount given by $2R_d-D_t$. The overlap ensures that the transmitted beam can still be detected by the acquisition sensor, if the jitter-induced excursion during the sensor integration time is smaller than $R_d-D_t/2$.  For larger excursions the remote transceiver cannot detect the beam, if the former is located in the hatched region and close to the middle between track $n$ and $n+1$, i.e. close to position $D_t/2$, and if the beam excursion is towards track $n-1$ in negative x-direction. Note that the impact on the detection probability is the same if the transceiver location jitters in the uncertainty plane instead of the scanning beam, but into the opposite direction. We use this relation in order to simplify the analyses.

We shall assume that the location of the remote transceiver is described by a random variable $X_t$ that is uniformly distributed between $[0,D_t/2]$. The assumption of uniformity is a good approximation for most cases, as $D_t/2$ is typically much smaller than the width $\sigma$ of the normal distribution defining the uncertainty of the transceiver location. All other cases, e.g. if the transceiver is located in between $[D_t/2,D_t]$, are covered by this assumption as well for reasons of symmetry and reciprocity.
The incident scanning beam is sampled by the acquisition sensor with a frequency $f_s$ that is the inverse of the integration time $T_{int}$. The integration time is set to half the time it it takes for the full width ($2\,R_d$) of the scanning beam to sweep across the detector (and not the full time) to avoid unnecessary accumulation of stray light, if the remote TC is located off beam-center and the sweep time is shorter.
When the sensor integrates the incident light,  it averages out the fluctuations occurring at frequencies larger than $f_s/2$. 
To be more precise, as the laser beam profile is convex at the beam waist, unbiased jitter around a transceiver location close to the detection radius actually leads to a higher average intensity than if the beam was remaining static so that the assumption that fluctuations are merely averaging out is erring on the conservative side.

Without loss of generality, we consider the case that the beam is sweeping across the transceiver location while it is moving along spiral track $n$. We also assume that the RMS jitter of the scanning beam for a sampling frequency $f_s$ is described by a random variable $X_n$ that  that is normally distributed with zero mean and a variance $\sigma_n^2$ and a PDF given by a Gaussian function $g(x_n,\sigma_n^2)$. As mentioned before, instead of displacing the beam, we equivalently assume the jitter superposes the positional coordinates of the transceiver in the uncertainty plane. Assuming the transceiver was not "hit" while the beam was sweeping along track $n$, the beam will later on move along spiral track $n+1$ and - considering the overlap between the beam profiles - have another chance of hitting the transceiver when it passes through the radial vector $R_{TC}$. Here again we assume that the RMS jitter of the scanning beam is described a random variable $X_{n+1}$ that is normally distributed with a variance $\sigma_n^2$. We now make the assumption that the two random variables $X_n$ and $X_{n+1}$ are uncorrelated, i.e. $\langle X_n X_{n+1}\rangle=0$. 

We now define an event $A$ to represent the failure of the scanning beam to "hit" the transceiver on the spiral passage along track $n$ and the event $B$ to represent the failure to hit it along track $n+1$.
From the discussion above and applying the definitions of Fig.\ref{fig:overlapping_beams}, we are able to define equations for the probabilities $P(A)$ and $P(B)$ for events A and B to occur:
\begin{eqnarray}
P(A) &=&\text{Prob}(X_t + X_n > R_d)\label{eq:P(A)}\\
P(B) &=&\text{Prob}(D_t- X_t - X_{n+1} > R_d)\label{eq:P(B)}
\end{eqnarray}
Note that the sign of $X_n$  and $X_{n+1}$ in the above equations does not matter because both are unbiased, independent, normal variates. However, for the sake of clarity, we followed the definition from above that the beam jitter is represented by fluctuations of the transceiver coordinates which are defined positive in the direction from inner to outer track. We note that there is a correlation between Eq. \ref{eq:P(A)} and Eq. \ref{eq:P(B)} so that the overall probability of hitting the transceiver on neither of the two tracks is not simply the product of $P(A)$ and $P(B)$ but depends on the conditional probability $P(B|A)$. In order to compute all relevant quantities we marginalise Eqs. \ref{eq:P(A)} and \ref{eq:P(B)} over the random variable $X_t$. Consequently, this allows expressing the overall probability $P(A\&B)$ as a product of two marginal probabilities and integrating over the marginal parameter $x_t$,  weighted by its probability density function $P(x_t)$. We then obtain
\begin{eqnarray*}
P(A)&=&\int dx_t P(A|X_t=x_t) P(x_t)\\
P(B)&=&\int dx_t P(B|X_t=x_t) P(x_t)\\
P(A\&B)&=&\int dx_t  P(A|X_t=x_t)P(B|X_t=x_t) P(x_t)
\end{eqnarray*}
These integrals can be readily computed as all associated parameters are now fully defined:
\begin{eqnarray}
P(A|X_t=x_t)&=&P(X_n>R_d-x_t)=\int_{R_d-x_t}^{\infty}dx_tg(x_n)\nonumber\\
&=& \frac{1}{2}\text{erfc}\left(\frac{R_d-x_t}{\sqrt{2}\sigma_n}\right)\label{eq:P_A_temp}
\end{eqnarray}
where $\text{erfc}(x)$ is the complementary error function. Integrating Eq.\ref{eq:P_A_temp} over the marginal parameter $x_t$ we find $P(A)$, and applying a similar approach we also find the probabilities $P(B)$ and $P(B|A)$. The results are summarized below
\begin{eqnarray}
P(A)&=&\frac{1}{D_t}\int_0^{\frac{D_t}{2}}dx_t\text{erfc}\left(\frac{R_d-x_t}{\sqrt{2}\sigma_n}\right)\label{eq:P_A}\\
P(B)&=&\frac{1}{D_t}\int_0^{\frac{D_t}{2}}dx_t\text{erfc}\left(\frac{x_t-D_t+R_d}{\sqrt{2}\sigma_n}\right)\label{eq:P_B}\\
P(A\&B)&=&\frac{1}{2D_t}\int_0^{\frac{D_t}{2}}\!\!\!\!dx_t\text{erfc}\left(\frac{R_d-x_t}{\sqrt{2}\sigma_n}\right)\!\text{erfc}\left(\frac{x_t-D_t+R_d}{\sqrt{2}\sigma_n}\right)\nonumber\\
&&\label{eq:P_A_and_B}
\end{eqnarray}
The conditional probability $P(B|A)$ that the remote transceiver is not detected on track $n+1$ given that it has also not been detected on track $n$ is defined by $P(B|A)=P(A\&B)/P(A)$. It is found as the ratio between Eq. \ref{eq:P_A_and_B} and Eq.\ref{eq:P_A}.
%%%%%%%%%%%%%%%%%%%%%%%%%%%%%%%%%%%%%%%%%%%%%%%%%%%%%%%%%%%%%%%%%%%%%%%%%%%%%
\subsection{Discussion of results}
\label{sec:jitter_discussion}
\begin{figure}
\centerline{\includegraphics[width=\columnwidth]{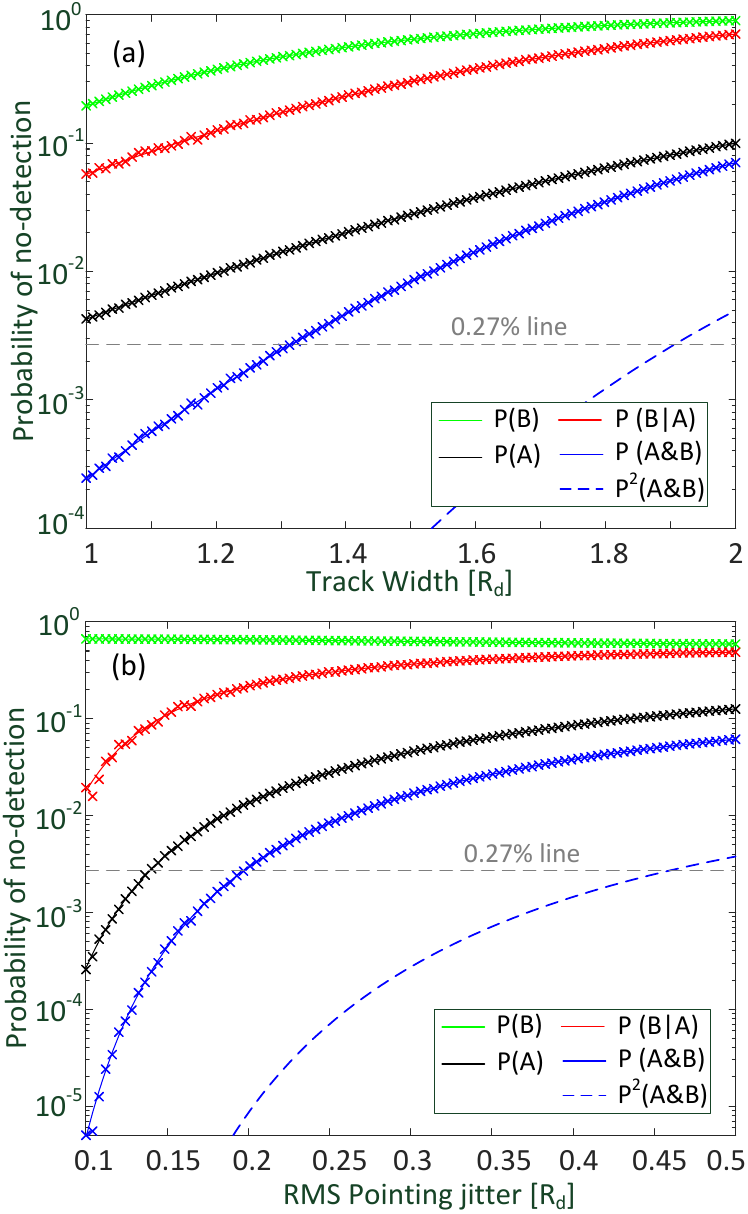}}
\caption  {From top to bottom: The probabilities $P(B)$, $P(B|A)$, $P(A)$, $P(A\&B)$,  $P^2(A\&B)$ are plotted against variations of the track width $D_t$ in (a) and against variations of the RMS beam jitter $\sigma_n$ in (b),  both expressed in units of detection radius $R_d$. Beam jitter $\sigma_n=0.25 R_d$ in (a) and track width $D_t=1.5 R_d$ in (b)\label{fig:prob_det}. }
\end{figure}

Figure \ref{fig:prob_det} plots the various probabilities against changes of the track width $D_t$ for a given $\sigma_n=0.25 R_d$ in (a) and against changes of the beam jitter RMS level $\sigma_n$ for a given $D_t=1.5 R_d$ in (b).
We see in (a) that $P(A)$, i.e. the probability of not-detecting the TC on track $n$ (black curve), is high at 10\% for $D_t=2 R_d$, where the beams do not overlap. It decreases steadily down to a value of 0.4\% - which is still higher than the "$3\sigma$ value" we try to achieve (grey dashed line) - for $D_t=R_d$, where the right half of a beam on one track fully overlaps with the left half of a beam on the other track. However, the overall probability of detecting the TC neither on track $n$ nor on track $n+1$ (blue solid curve) is decreasing below the $3\sigma$ threshold for $D_t=1.3 R_d$, where the two halves of the beams have 70\% overlap.
This indicates that there is a significant reduction of the failure rate due to the second passage on track $n+1$, which is confirmed by the conditional probability $P(B|A)$ that the TC is not detected on track $n+1$, given that it was missed before on track $n$ (red curve). $P(B|A)$ is as low as 6\% for $D_t=1.0 R_d$, indicating that in 94\% of the cases the TC is detected on track $n+1$ after it was missed before on track $n$. This clearly demonstrates the crucial role of the neighbouring track as a "safety net" to detect the vast majority of cases that were missed on the track before.
The blue dashed line denotes the probability that the TC is not detected when a "dual scan" is performed for the link acquisition (see section 2\ref{sec:dual_scan}). When the dual scan method is used, both transceiver scans must fail for the overall acquisition to fail. Therefore, the overall failure probability is given by the square of $P(A\&B)$, considering that the beam jitters of TC1 and TC2 are not correlated. This makes the dual scan method very robust against failure, which is an additional benefit on top of the faster acquisition times. 
\\The beam jitter is varied in Fig. \ref{fig:prob_det}(b) which shows that the total probability of failure $P(A\&B)$ is 6\% for large jitter with $\sigma_n=0.5\,R_d$, and steadily decreases to a level below the 0.27\% threshold line for a jitter with $\sigma_n=0.2\,R_d$. Here again the additional "safety net" of catching a TC during track $n+1$ after missing it on track $n$ can significantly increase the success rate, as demonstrated by the curve for the conditional probability $P(B|A)$. 
In order to validate the analytical derivations starting from Eqs. \ref{eq:P(A)} and \ref{eq:P(B)}, we performed Monte Carlo simulations the results of which are given by the crosses in Fig. \ref{fig:prob_det}(a) and (b) and are found to be in good agreement with the analytical predictions.
\\It should be noted that our model accounts for an "average displacement'' of the scanning beam while it is sweeping past the location of the transceiver in the uncertainty plane, which is representative of a detector integrating the received power from the scanning beam during the passage time $T_{pass}=2R_d/\gamma$. 
For jitter frequencies $f$ whose spatial length $L_{jitter}=\gamma/f$ is significantly larger than the detection radius $R_d$, the amplitude of the scanning beam hardly changes during $T_{pass}$.
Therefore, if an integration time $T_{int}$ smaller than $T_{pass}$ is chosen, the individual integration windows are correlated and the predictions of Eq. \ref{eq:P_A_and_B} are found to be in good agreement with the results of Monte Carlo simulations of the acquisition dynamics, irrespective of the integration time (with the parameters defined in section 2, this was found to hold for spectra limited to frequencies $f< 0.1\,\text{Hz}$). 
If the jitter spectrum extends to higher frequencies, the integration windows become increasingly uncorrelated, and -in the limit of a spectrum extending beyond half the sampling frequency- become fully uncorrelated. 
In this case, the probabilities of the individual integration windows must be multiplied in order to obtain the overall probability of failure, which requires adjusting Eq.\ref{eq:P_A_and_B} accordingly. This is the “uncorrelated” limit assumed in the derivation of the model in \cite{friederichs2016vibration}. However, the latter model does not account for a random distribution of the TC in the uncertainty plane and for correlations between tracks which we found to be important in our discussion above.
\\We shall now look at an example for how to use the results of this section in the definition of critical system parameters. After choosing a track width $D_t=1.3 R_d$ and a tolerable jitter of $\sigma_n=0.25 R_d$, we come back to the discussion of spiral parameters in section 2\ref{sec:ProbDistr}. We now know how much overlap is required between the tracks and must therefore reduce the track width from 270 $\mu$rad  (2$R_d$) to 176 $\mu$rad (1.3 $R_d$), while leaving the scanning speed as before at one detection radius $R_d$ per second, which increases the mean search time from $131\,s$ to $202\,s$. We also have to ensure that the RMS jitter does not exceed $\sigma_n=34\, \mu\text{rad}$ ($0.25\,R_d$). The beam jitter is dominated by the low frequency fluctuations of the SC pointing error which in turn depends on the noise of the star cameras onboard. Recent analyses of data of the original GRACE mission\cite{inacio2015analysis} have shown the noise of the camera in the cross bore-sight directions to be $23\,\mu\text{rad}\,(1\sigma)$ for a sampling rate of $1\,\text{Hz}$, which is well below the $\sigma_n=34\, \mu\text{rad}$ we can tolerate for the chosen search parameters.   

%%%%%%%%%%%%%%%%%%%%%%%%%%%%%%%%%%%%%%%
\section{Conclusion}
We have derived simple analytical expressions for the probability density functions (PDFs) of the time it takes to acquire optical links using a dedicated wide-field acquisition sensor and following either a single scan or a dual scan spiral search method.
These results were used to derive expressions for the PDFs of acquisition search times for link chains, where we investigated the cases that the links are either established sequentially or in parallel. A total of 4 different methods was introduced and compared in performance, where all analytical predictions were shown to agree very well with the results of Monte Carlo simulations. We also studied the impact of jitter of the scanning beam on the probability of failing to acquire the link on the basis of a derived analytical model that predicts the failure rate for given track width, detection radius and jitter magnitude.

The analyses show that it is very advantageous to use the dual scan method to acquire the individual links between spacecraft, as this method is twice as fast and much more robust against beam jitter than the single scan method. On the downside, this requires an acquisition sensor with a FoV twice as large as the one needed for a single scan, which may not always be economical. When acquiring a constellation of 3 satellites, parallel acquisition of all links is on average 1.6 times faster than sequential acquisition. However, this gain in speed must be traded against higher operational complexity, because each spacecraft must engage in two links at the same time. We also find that the probability of acquisition failure due to beam jitter can be reduced by a factor $\sim 300$ if the spiral pitch $D_t$ is reduced from $2.0\,R_d$ to $1.0\,R_d$, which is important to consider in the choice of acquisition parameters.
%%%%%%%%%%%%%%%%
\begin{backmatter}
\bmsection{Acknowledgments} The author gratefully acknowledges fruitful discussions with Jacopo Aurigi, Simon Delchambre, Tobias Ziegler,  Christian Greve and R\"udiger Gerndt (Airbus).'

\bmsection{Disclosures} The author declares no conflicts of interest.

\bmsection{Data availability} The data underlying the results presented in this paper were generated by equations listed in the text. 

\end{backmatter}

%%%%% References %%%%%


\begin{thebibliography}{10}
\newcommand{\enquote}[1]{``#1''}

\bibitem{fields2009nfire}
R.~Fields, C.~Lunde, R.~Wong, J.~Wicker, D.~Kozlowski, J.~Jordan, B.~Hansen,
  G.~Muehlnikel, W.~Scheel, U.~Sterr \emph{et~al.},
  \enquote{Nfire-to-terrasar-x laser communication results: satellite pointing,
  disturbances, and other attributes consistent with successful performance,}
  in \emph{Sensors and Systems for Space Applications III,}  vol. 7330
  (International Society for Optics and Photonics, 2009), p. 73300Q.

\bibitem{smutny2008orbit}
B.~Smutny, R.~Lange, H.~K{\"a}mpfner, D.~Dallmann, G.~M{\"u}hlnikel,
  M.~Reinhardt, K.~Saucke, U.~Sterr, B.~Wandernoth, and R.~Czichy,
  \enquote{In-orbit verification of optical inter-satellite communication links
  based on homodyne bpsk,} in \emph{Free-Space Laser Communication Technologies
  XX,}  vol. 6877 (International Society for Optics and Photonics, 2008), p.
  687702.

\bibitem{hauschildt2019global}
H.~Hauschildt, N.~le~Gallou, S.~Mezzasoma, H.~L. Moeller, J.~P. Armengol,
  M.~Witting, J.~Herrmann, and C.~Carmona, \enquote{Global
  quasi-real-time-services back to europe: Edrs global,} in \emph{International
  Conference on Space Optics—ICSO 2018,}  vol. 11180 (International Society
  for Optics and Photonics, 2019), p. 111800X.

\bibitem{heinzel2017laser}
G.~Heinzel, B.~Sheard, N.~Brause, K.~Danzmann, M.~Dehne, O.~Gerberding,
  C.~Mahrdt, V.~M{\"u}ller, D.~Sch{\"u}tze, G.~Stede \emph{et~al.},
  \enquote{Laser ranging interferometer for grace follow-on,} in
  \emph{International Conference on Space Optics—ICSO 2012,}  vol. 10564
  (International Society for Optics and Photonics, 2017), p. 1056420.

\bibitem{schutze2014laser}
D.~Sch{\"u}tze, G.~Stede, V.~M{\"u}ller, O.~Gerberding, T.~Bandikova, B.~S.
  Sheard, G.~Heinzel, and K.~Danzmann, \enquote{Laser beam steering for grace
  follow-on intersatellite interferometry,} {\protect\JournalTitle{Optics
  express}} \textbf{22}, 24117--24132 (2014).

\bibitem{abich2019orbit}
K.~Abich, A.~Abramovici, B.~Amparan, A.~Baatzsch, B.~B. Okihiro, D.~C. Barr,
  M.~P. Bize, C.~Bogan, C.~Braxmaier, M.~J. Burke \emph{et~al.},
  \enquote{In-orbit performance of the grace follow-on laser ranging
  interferometer,} {\protect\JournalTitle{Physical review letters}}
  \textbf{123}, 031101 (2019).

\bibitem{danzmann2003lisa}
K.~Danzmann and L.~S. Team, \enquote{Lisa—an esa cornerstone mission for the
  detection and observation of gravitational waves,}
  {\protect\JournalTitle{Advances in Space Research}} \textbf{32}, 1233--1242
  (2003).

\bibitem{luo2020taiji}
Z.~Luo, Y.~Wang, Y.~Wu, W.~Hu, and G.~Jin, \enquote{The taiji program: A
  concise overview,} {\protect\JournalTitle{Progress of Theoretical and
  Experimental Physics}}, 2050--3911,  (2020).

\bibitem{sato2017status}
S.~Sato, S.~Kawamura, M.~Ando, T.~Nakamura, K.~Tsubono, A.~Araya, I.~Funaki,
  K.~Ioka, N.~Kanda, S.~Moriwaki \emph{et~al.}, \enquote{The status of decigo,}
  in \emph{Journal of Physics: Conference Series,}  vol. 840 (IOP Publishing,
  2017), p. 012010.

\bibitem{benzi2016optical}
E.~Benzi, D.~C. Troendle, I.~Shurmer, M.~James, M.~Lutzer, and S.~Kuhlmann,
  \enquote{Optical inter-satellite communication: the alphasat and sentinel-1a
  in-orbit experience,} in \emph{14th International Conference on Space
  Operations,}  (2016), p. 2389.

\bibitem{cirillo2009control}
F.~Cirillo and P.~F. Gath, \enquote{Control system design for the constellation
  acquisition phase of the lisa mission,} in \emph{Journal of Physics:
  Conference Series,}  vol. 154 (IOP Publishing, 2009), p. 012014.

\bibitem{mahrdt2014laser}
C.~Mahrdt, \enquote{Laser link acquisition for the grace follow-on laser
  ranging interferometer,} Ph.D. thesis, Hannover: Gottfried Wilhelm Leibniz
  Universit{\"a}t Hannover (2014).

\bibitem{wuchenich2014laser}
D.~M. Wuchenich, C.~Mahrdt, B.~S. Sheard, S.~P. Francis, R.~E. Spero,
  J.~Miller, C.~M. Mow-Lowry, R.~L. Ward, W.~M. Klipstein, G.~Heinzel
  \emph{et~al.}, \enquote{Laser link acquisition demonstration for the grace
  follow-on mission,} {\protect\JournalTitle{Optics express}} \textbf{22},
  11351--11366 (2014).

\bibitem{scheinfeild2000}
M.~Scheinfeild, N.~S. Kopeika, and R.~Melamed, \enquote{{Acquisition system for
  microsatellites laser communication in space},} in \emph{Free-Space Laser
  Communication Technologies XII,}  vol. 3932 G.~S. Mecherle, ed.,
  International Society for Optics and Photonics (SPIE, 2000), pp. 166 -- 175.

\bibitem{scheinfeild2001}
M.~Scheinfeild, N.~S. Kopeika, and A.~Shlomi, \enquote{{Acquisition time
  calculation and influence of vibrations for microsatellite laser
  communication in space},} in \emph{Acquisition, Tracking, and Pointing XV,}
  vol. 4365 M.~K. Masten and L.~A. Stockum, eds., International Society for
  Optics and Photonics (SPIE, 2001), pp. 195 -- 205.

\bibitem{hindman2004}
C.~{Hindman} and L.~{Robertson}, \enquote{Beaconless satellite laser
  acquisition - modeling and feasability,} in \emph{IEEE MILCOM 2004. Military
  Communications Conference, 2004.},  vol.~1 (2004), pp. 41--47 Vol. 1.

\bibitem{li2011analytical}
X.~Li, S.~Yu, J.~Ma, and L.~Tan, \enquote{Analytical expression and
  optimization of spatial acquisition for intersatellite optical
  communications,} {\protect\JournalTitle{Optics express}} \textbf{19},
  2381--2390 (2011).

\bibitem{ma2021satellite}
J.~Ma, G.~Lu, L.~Tan, S.~Yu, Y.~Fu, and F.~Li, \enquote{Satellite platform
  vibration influence on acquisition system for intersatellite optical
  communications,} {\protect\JournalTitle{Optics \& Laser Technology}}
  \textbf{138}, 106874 (2021).

\bibitem{friederichs2016vibration}
L.~Friederichs, U.~Sterr, and D.~Dallmann, \enquote{Vibration influence on hit
  probability during beaconless spatial acquisition,}
  {\protect\JournalTitle{Journal of Lightwave Technology}} \textbf{34},
  2500--2509 (2016).

\bibitem{luo2017possible}
Z.~Luo, Q.~Wang, C.~Mahrdt, A.~Goerth, and G.~Heinzel, \enquote{Possible
  alternative acquisition scheme for the gravity recovery and climate
  experiment follow-on-type mission,} {\protect\JournalTitle{Applied Optics}}
  \textbf{56}, 1495--1500 (2017).

\bibitem{kawamura2019space}
S.~Kawamura, T.~Nakamura, M.~Ando, N.~Seto, T.~Akutsu, I.~Funaki, K.~Ioka,
  N.~Kanda, I.~Kawano, M.~Musha \emph{et~al.}, \enquote{Space
  gravitational-wave antennas decigo and b-decigo,}
  {\protect\JournalTitle{International Journal of Modern Physics D}}
  \textbf{28}, 1845001 (2019).

\bibitem{inacio2015analysis}
P.~In{\'a}cio, P.~Ditmar, R.~Klees, and H.~H. Farahani, \enquote{Analysis of
  star camera errors in grace data and their impact on monthly gravity field
  models,} {\protect\JournalTitle{Journal of Geodesy}} \textbf{89}, 551--571
  (2015).

\end{thebibliography}
\end{document}